\documentclass{article}
\usepackage{srcltx}
\usepackage[T2A]{fontenc}
\usepackage[utf8]{inputenc}
\usepackage[english]{babel}
\usepackage{graphicx}
\usepackage{amsmath,amsfonts,amssymb,amsthm,mathtools}
\usepackage{booktabs}
\begin{document}
\date{}

\title{ Neutron Stars in a Uniform Density Approximation}

\author{G. S. Bisnovatyi-Kogan \footnote{{Space Research Institute, Russian Academy of Sciences, Moscow, 117997 Russia,};
{National Research Nuclear University MEPhI, Moscow, 115409 Russia };
{Moscow Institute of Physics and Technology, Dolgoprudny, Moscow oblast, 141701 Russia.}},
E. A. Patraman \footnote{{Space Research Institute, Russian Academy of Sciences, Moscow, 117997 Russia,};
{Moscow Institute of Physics and Technology, Dolgoprudny, Moscow oblast, 141701 Russia}}}

\maketitle

\begin{abstract}
      Models of neutron stars are considered in the case of a uniform density distribution. A universal
algebraic equation, valid for any equation of state, is obtained. This equation allows one to find the approximate mass of a star of a given density without resorting to the integration of differential equations. The solutions presented in the paper for various equations of state, including more realistic ones, differ from the exact
solutions obtained by the numerical integration of differential equations by at most $\sim 20 \%$.
\end{abstract}

\section{INTRODUCTION}
When studying the structure of white dwarfs, it was
discovered that their equilibrium is possible only for
masses not exceeding a certain limit, known as Chandrasekhar’s. For a carbon-oxygen chemical composition, with two baryons per electron, $\mu_e=2$ , this limit
is  $\approx 1.46\, M_\odot$. The first conclusion about the existence
of an upper mass limit for cold stars, the equilibrium of
which is maintained by the pressure of degenerate
electrons, was made by Stoner \cite{stoner30}, who considered the
model of a white dwarf of uniform density. He generalized the consideration of the pressure of degenerate
electrons, made in \cite{fowler26,frenkel28}, to the case of high density
under conditions of ultra-relativistic degeneracy, in
which the equation of state for cold matter takes the
form \cite{bk89}

\begin{equation}
\label{eq1}
P(\rho)=\frac{(3\pi^2)^{1/3}}{4}\frac{\hbar c}{(\mu_e m_u)^{4/3}}\rho^{4/3}= K\,\rho^{4/3}.
\end{equation}
Here $\mu_e$, is the number of baryons per electron, $m_u$ is
an atomic mass unit equal to 1/12 of the mass of the
$^{12}C$ isotope. According to Emden’s theory, the mass of
a polytropic star , corresponding to $\gamma=1+1/n =4/3,\,\, n=3$
, does not depend on density and is uniquely
determined by the parameter $K$ as \cite{bk89}

\begin{equation}
\label{eq2}
M_p=4\pi\left(\frac{K}{\pi G}\right)^{3/2}\, M_3, \quad M_3=2.01824.
\end{equation}
Using (\ref{eq2}), Chandrasekhar \cite{chandra31} and Landau \cite{landau32}, independently and almost simultaneously, obtained for the equation of state (\ref{eq1}) the limiting mass of a white dwarf
in the form

\begin{equation}
\label{eq3}
M_{wd}=\frac{\sqrt{3\pi}}{2}\left(\frac{\hbar c}{G}\right)^{3/2} \frac{M_3}{(\mu_e m_u)^2}=\frac{5.83}{\mu_e^2} M_\odot.
\end{equation}
To determine the maximum mass of observed white
dwarfs, Chandrasekhar \cite{chandra31}, following Stoner \cite{stoner30}, used $\mu_e=2.5$
 and obtained $M_{wd}=0.933 M_\odot$. This refined
the value of $M_{wd}=1.1 M_\odot$ calculated by Stoner using
the same value of $\mu_e=2.5$  in the uniform density
model. From the theory of stellar evolution, as well as
from observations, it follows that almost all white
dwarfs consist of a mixture of carbon $^{12}C$ and oxygen
$^{16}O$ for which $\mu_e=2$ \cite{shatzman} and $M_{wd}=1.46 M_\odot$. In [\cite{landau32}],
for the first time, a realistic value for the mass limit of
a white dwarf was obtained, which deserves to be
referred to as the Stoner–Chandrasekhar–Landau
limit. Here, we used refined modern values for all constants \cite{bkp}, which led to a difference of several percent
from the values given in the original works.

  In this paper, we construct approximate models of
cold neutron stars of arbitrary mass assuming a uniform density distribution by using algebraic equations derived from the general theory of relativity
(GTR). Within this model, all results, including the
values of the limiting masses of neutron stars (NSs),
are obtained analytically, from algebraic equations we
derived, for any equations of state.

\section{UNIFORM-DENSITY NEUTRON STARS
}
To construct realistic models of NSs, it is necessary
to use general relativity, since the gravitational potential reaches tenths of $c^2$ and the NS radius $R_{ns}$ is only a
few gravitational radii $r_g=2GM/c^2$. The models of
non-rotating NSs are constructed using a
Schwarzschild-type metric \cite{lltp,zn71,oveq}

\begin{eqnarray}
\label{eq19}
ds^2=e^{\nu(r)}c^2dt^2-e^{\lambda(r)}dr^2-r^2(d\theta^2+\sin^2\theta d\phi^2),\quad e^{-\lambda}=1-\frac{2Gm}{rc^2}, \qquad\\
e^{\nu} = \left(1 - \frac{2 G m}{r c^2}\right)exp\left[ \int_0^{P(r)}\frac{2dP}{P+\rho(P)}\right],\qquad \qquad\qquad \nonumber
\end{eqnarray}
where

 \begin{eqnarray}
\label{eq20}
m(r)=4\pi\int_0^r \rho r'^2 dr', \quad M=m(R_{ns}), \quad \rho=\rho_0\left(1+\frac{E_T}{c^2}\right).
\end{eqnarray}
In general relativity, they consider the total density of
matter $\rho$, which includes the total rest energy density  $\rho_0$
 and internal energy $E_T$. The mass $M$ is the total
gravitating mass, which includes the gravitational
energy; therefore, the total energy of an NS is ${\cal E}=M c^2$
 and the total energy $e$ inside a radius $r$ is $e(r)=m(r) c^2$
. From the general relativity equations it
follows that the baryon density $n$ is related to the number of baryons $f(r)$ inside the radius $r$, as \cite{bk89}

 \begin{eqnarray}
\label{eq21}
f(r)= 4\pi\int_0^r n\, r^2\left(1-\frac{2Gm}{c^2 r}\right)^{-1/2} dr,\quad N=f(R_{ns}), \quad \rho_0=n\,m_u, \quad M_0=N\,m_u,
\end{eqnarray}
where $M_0$ - is the baryonic rest mass in the NS. For a
uniform-mass NS, in which $n$, $\rho_0$, $E_T$ , and $\rho$ do not
depend on the radius, the integral in expression (\ref{eq21}) is
calculated analytically, which leads to the expression
\cite{ryzhik} 

 \begin{eqnarray}
\label{eq22}
f(r)= \frac{2\pi n}{D^{3/2}}\left[\sin^{-1}(r\sqrt{D})-r\sqrt{D}\sqrt{1-r^2D}\right], \quad D=\frac{8\pi \rho G}{3c^2}.
\end{eqnarray}
To relate the NS radius $R$ to the density $\rho$ , it is necessary to find the extremum of the function $M(\rho_0)$ at a fixed value of the number of baryons  in the star, $N$ ,
which are written as

 \begin{eqnarray}
\label{eq23}
M=\frac{4\pi}{3}\rho R^3=\frac{4\pi}{3 c^2}\rho_0(c^2+E_T) R^3, \qquad
N= \frac{2\pi \rho_0}{m_u D^{3/2}}\left[\sin^{-1}(R\sqrt{D})-R\sqrt{D}\sqrt{1-R^2D}\right].
\end{eqnarray}
If we are not given a specific density distribution function, then, applying the variational principle and using
expressions (\ref{eq20}),(\ref{eq21}) to obtain a differential equation and find the density distribution function \cite{bk89}, we
find the equilibrium equation for a star in general relativity, obtained by Oppenheimer and Volkoff \cite{oveq}. As
shown in \cite{bk89}, p. 407, for this it is necessary to find the
variation of the total energy (mass) as a functional of
the variation of the radius $\delta r(\nu)$ , where $\nu \equiv f(r)$,
obtained in expression (\ref{eq21}).

\subsection{Calculation of Differentials}
Taking the differentials of both functions in (\ref{eq23}), we
get

\begin{eqnarray}
\label{eq24}
dM=\frac{4\pi}{3 c^2}[3R^2 \rho_0(c^2+E_T) dR+ (c^2+E_T) R^3d\rho_0+\rho_0 R^3 dE_T],\qquad\quad \\
\label{eq24a}
dN= \frac{2\pi}{m_u}\left[\sin^{-1}(R\sqrt{D})-R\sqrt{D}\sqrt{1-R^2D}\right]
\left(\frac{d\rho_0}{D^{3/2}}-\frac{3\rho_0}{2 D^{5/2}}dD \right)
+\frac{2\pi \rho_0}{m_u D^{3/2}}\qquad\\
\times\left[\frac{R dD+2{D} dR}{2\sqrt{D}\sqrt{1-R^2 D}} - \sqrt{1-R^2D}\left(\sqrt{D}dR+\frac{R}{2\sqrt{D}}dD\right)
+R\sqrt{D}\frac{R^2 dD+2D RdR}{2\sqrt{1-R^2 D}}
\right] .\nonumber
\end{eqnarray}
From thermodynamic relations, we obtain \cite{llsp}

\begin{eqnarray}
\label{eq25}
dE_T=\left(\frac{P}{\rho_0^2}\right)d\rho_0, \qquad
dD=\frac{8\pi G}{3 c^2}d\rho= \frac{8\pi G}{3 c^2}(1+\frac{E_T}{c^2})d\rho_0 \\ + \frac{8\pi G}{3 c^2}\frac{P}{\rho_0 c^2}d\rho_0=\frac{8\pi G}{3c^2}\left(1+\frac{E_T}{c^2}+\frac{P}{c^2
\rho_0}\right)d\rho_0 . \nonumber
\end{eqnarray}
Given (\ref{eq25}), we obtain from (\ref{eq24}) and (\ref{eq24a}) expressions for
the differentials $dM(d\rho_0, dR)$ and $dN(d\rho_0, dR)$. Considering the variations for a conserved number of baryons, with $dN=0$, we obtain from the last expression
a relation between the differentials $dR$ and $d\rho_0$. The equilibrium of a uniform-density star corresponds to
the extremum of the function $M(\rho_0)$, in which
\begin{equation}
\label{eq30}
\frac{dM}{d\rho_0}=0.
\end{equation}
Then, from the expressions for differentials, we obtain
a relation between the radius $R$ of a uniform-density
star in general relativity and the total density $\rho=\rho_0\left(1+\frac{E_T}{c^2}\right)$ in the form

\begin{eqnarray}
\label{eq31}
 \frac{\sqrt{1-R^2D}}{2R^3\sqrt{D^3}}
\left[R\sqrt{D}\sqrt{1-R^2D}-\sin^{-1}(R\sqrt{D})\right] \qquad\qquad
\nonumber\\
+\frac{2\pi G}{c^2}\,\frac{\sqrt{1-R^2D}}{2R^2 D^2}\,\,
\frac{\rho c^2+P}{3 c^2}
\left[\frac{\sin^{-1}(R\sqrt{D})}{R\sqrt{D}}-
\frac{1-\frac{1}{3}R^2D}{\sqrt{1-R^2D}}\right]+\frac{\rho c^2+P}{3\rho c^2}=0
\end{eqnarray}
Relation (\ref{eq31}) holds for any equations of state, allowing
one to approximately find the mass of a star of a given
density from an algebraic equation, without resorting
to the integration of differential equations.

    Expressions for the differentials $dM$ and $dN$ can be
obtained using integral formulas for these quantities,
following from (\ref{eq20}) and (\ref{eq21}), in the form   

\begin{eqnarray}
\label{eq32}
M=4\pi\int_0^{R} \rho r^2 dr, \quad N= 4\pi\int_0^{R} n\, r^2\left(1-\frac{2Gm}{c^2 r}\right)^{-1/2} dr,\quad m=\frac{4\pi}{3}\rho r^3, \quad \rho_0=m_u\,n.
\end{eqnarray}
The calculation of $dM$ using the integral expression in
(\ref{eq20}) does not lead to simplifications, but the calculation
of $dN$ using the integral expression is greatly simplified. The integrals arising after using (\ref{eq32}) are calculated analytically, which results in the expression

\begin{eqnarray}
\label{eq36}
dN=\frac{4\pi \rho_0}{m_u}\frac{R^2 dR}{\sqrt{1-R^2 D}} + \frac{2\pi \rho_0}{m_u}\frac{R^3}{\sqrt{1-R^2 D}}\frac{dD}{D} \nonumber\\
+\frac{2\pi \rho_0}{m_u D^{3/2}}\left(\frac{d\rho_0}{\rho_0}-\frac{3}{2}\frac{dD}{D}\right) [\,\sin^{-1}(R\sqrt{D})-R\sqrt{D}\sqrt{1-R^2D}\,]
\end{eqnarray}
which relates $dR$ and $d\rho_0$ and allows the differential $dM$
 to be written in the form
 \begin{eqnarray}
\label{eq39}
dM=4\pi\rho R^3\left[-\frac{1}{6}\frac{dD}{D}
-\frac{\sqrt{1-R^2D}}{2 D^{3/2}R^3}
[\,\sin^{-1}(R\sqrt{D})-R\sqrt{D}\sqrt{1-R^2D}\,]\left(\frac{d\rho_0}{\rho_0}-\frac{3}{2}\frac{dD}{D}\right)\right].\nonumber\\
\end{eqnarray}
As a result, from (\ref{eq39}), considering (\ref{eq30}), we obtain an
algebraic equation that determines the equilibrium of
a uniform-density star, identical to (\ref{eq31}):

\begin{eqnarray}
\label{eq40}
\frac{1}{6}\frac{d\rho}{d\rho_0}\frac{\rho_0}{\rho}
+\frac{\sqrt{1-x^2}}{2 x^3}
[\,\sin^{-1}(x)-x\sqrt{1-x^2}\,]\left(1-\frac{3}{2}\frac{d\rho}{d\rho_0}\frac{\rho_0}{\rho}\right)=0, \qquad
\frac{d\rho}{d\rho_0}\frac{\rho_0}{\rho}
=\frac{1+\frac{E_T}{c^2}+\frac{P}{c^2\rho_0}}{1+\frac{E_T}{c^2}}
\end{eqnarray}
For convenience, here we use the designations

\begin{eqnarray}
\label{eq41}
x=R\sqrt{D}=\sqrt{\frac{R_g}{R}}, \qquad R_g=\frac{2 GM}{c^2}, \qquad \frac{dD}{D} = \frac{d\rho}{\rho} =\frac{1+\frac{E_T}{c^2}+\frac{P}{c^2\rho_0}}{1+\frac{E_T}{c^2}}\frac{d\rho_0}{\rho_0}   
\end{eqnarray}
Algebraic equation (\ref{eq40}) is written in the form

\begin{eqnarray}
\label{eq42}
\frac{d\rho}{d\rho_0}\frac{\rho_0}{\rho}\left[\frac{1}{6}-\frac{3}{2}\frac{\sqrt{1-x^2}}{2 x^3}
(\sin^{-1}{x}-x\sqrt{1-x^2})\right]
+\frac{\sqrt{1-x^2}}{2 x^3}(\,\sin^{-1}x-x\sqrt{1-x^2})=0.
\end{eqnarray}
Given the expression for $\frac{d\rho}{d\rho_0}\frac{\rho_0}{\rho}$ in (\ref{eq40}), we obtain from (\ref{eq42})
a relation for pressure in the form

\begin{eqnarray}
\label{eq43}
\frac{P}{\rho c^2}= \frac{\Phi_0(x)}{\Phi_1(x)}.
\end{eqnarray}
where

\begin{eqnarray}
\label{eq43b}
\Phi_0(x)= 2 x^3-3\sqrt{1-x^2}(\,\sin^{-1}x-x\sqrt{1-x^2})=
x(3-x^2)-3\sqrt{1-x^2}\sin^{-1}x  ,\\
\Phi_1(x)= -2x^3+9\sqrt{1-x^2}(\sin^{-1}{x}-x\sqrt{1-x^2})=
 x(7x^2-9)+9 \sqrt{1-x^2}\sin^{-1}x . \nonumber
\end{eqnarray}

\section{POST-NEWTONIAN ASYMPTOTICS, $\bf{x^2\ll 1}$}
In the limit of weak gravity, at $x\ll 1$ in (\ref{eq41}), we
obtain from (\ref{eq43}) in the post-Newtonian approximation the following equation:

\begin{eqnarray}
\label{eq44}
\frac{P}{\rho c^2}=\frac{x^2}{10}\left(1+\frac{61}{70}x^2\right)
\end{eqnarray}
where
\begin{eqnarray}
 x^2=\frac{R_g}{R}=\frac{8\pi G}{3c^2}\rho R^2=
\frac{2G}{c^2}\left(\frac{4\pi\rho}{3}\right)^{1/3} M^{2/3}.\nonumber
\end{eqnarray}
Using the expansions from \cite{ryzhik}, we get the following
relations:

\begin{eqnarray}
\label{eq46}
 N= \frac{2\pi \rho_0}{m_u D^{3/2}}\left[\sin^{-1}(R\sqrt{D})-R\sqrt{D}\sqrt{1-R^2D}\right]\approx \frac{4\pi \rho_0 x^3}{3m_u D^{3/2}}
\left(1+\frac{3}{10} x^2 \right),\qquad\quad
\nonumber\\  \quad N_0=\frac{4\pi \rho_0 R_0^3}{3m_u}  \qquad x^2=D R^2=\frac{2 G M}{c^2 R},\quad R_g=\frac{2GM}{c^2}, \quad R_{g0}=\frac{2GM_0}{c^2}.\qquad\qquad
\end{eqnarray}
The index “0” denotes Newtonian quantities. From
the condition of conservation of the number of baryons
when the radius of the star changes, we relate the
radius $R$ in general relativity to the Newtonian radius $R_0$
in the post-Newtonian limit $x^2 \ll 1$ in the form

\begin{eqnarray}
\label{eq47}
N=N_0,\qquad
R^3\left(1+\frac{3}{10} x^2 \right) \approx R_0^3.
\end{eqnarray}
Using (\ref{eq46}) and (\ref{eq47}), we obtain the post-Newtonian
relation between the mass values in general relativity
and in the Newtonian metric in the form

\begin{eqnarray}
\label{eq48}
M=M_0 \left(1+\frac{E_T}{c^2}-\frac{3}{10}\frac{R_{g0}}{R_0}\right).
\end{eqnarray}
We write the first post-Newtonian relation in (\ref{eq44}) as

\begin{eqnarray}
\label{eq49}
\frac{P}{\rho_0^{4/3}}=\frac{1}{5}\,\frac{GM}{\rho_0^{4/3}}\,
\frac{\rho}{R}\left(1+\frac{61}{70}\frac{R_{g0}}{R_0}\right).
\end{eqnarray}
Hence, considering the post-Newtonian relation
between Newtonian and relativistic quantities in (\ref{eq47})
and (\ref{eq48}), we obtain a post-Newtonian algebraic equilibrium
equation for a uniform-density star in the form

 \begin{eqnarray}
\label{eq50}
\frac{P}{\rho_0^{4/3}}=\frac{1}{5}\,\frac{GM_0}{\rho_0^{1/3}R_0}\,
\left(1+\frac{2E_T}{c^2}+\frac{47}{70}\frac{R_{g0}}{R_0}\right),\qquad \frac{R_{g0}}{R_0}=\frac{2}{c^2}\left(\frac{4\pi}{3}\right)^{1/3} G M_0^{2/3} \rho_0^{1/3}
\end{eqnarray}
In the small second term in (\ref{eq50}), we use the equality of
gravitational and thermal energies in a Newtonian
polytropic star with an exponent 4/3 in the form \cite{bkp}

 \begin{eqnarray}
\label{eq51}
P=K \,\rho_0^{4/3}, \quad \gamma=\frac{d\ln P}{d\ln \rho_0}=\frac{4}{3}, \qquad E_T=\frac{3}{5}\left(\frac{4\pi}{3}\right)^{1/3}G M_0^{2/3}\rho_0^{1/3}.
\end{eqnarray}
Given (\ref{eq51}), the equilibrium equation (\ref{eq50}) is transformed to

 \begin{eqnarray}
\label{eq52}
\frac{P}{\rho_0^{4/3}}=\frac{1}{5}\,\frac{GM_0}{\rho_0^{1/3}R_0}\,
\left[1+\frac{89}{35 c^2}\left(\frac{4\pi}{3}\right)^{1/3} G M_0^{2/3}
\rho_0^{1/3}\right].
\end{eqnarray}
For the equation of state (\ref{eq51}), the equation of equilibrium
of a uniform-density star will be written as

 \begin{eqnarray}
\label{eq53}
K=\frac{1}{5}\,\left(\frac{4\pi}{3}\right)^{1/3} GM_0^{2/3}\,
\left[1+\frac{89}{35 c^2}\left(\frac{4\pi}{3}\right)^{1/3} G M_0^{2/3}
\rho_0^{1/3}\right]
\end{eqnarray}
This equation is identical to the corresponding
equation from \cite{bkp}, based on the post-Newtonian
expansions given in \cite{zn71}.

\section{UNIFORM DENSITY MODELS
AT VERY HIGH DENSITIES}
\begin{figure}[h]
\begin{center}
\includegraphics[width=0.7\linewidth]{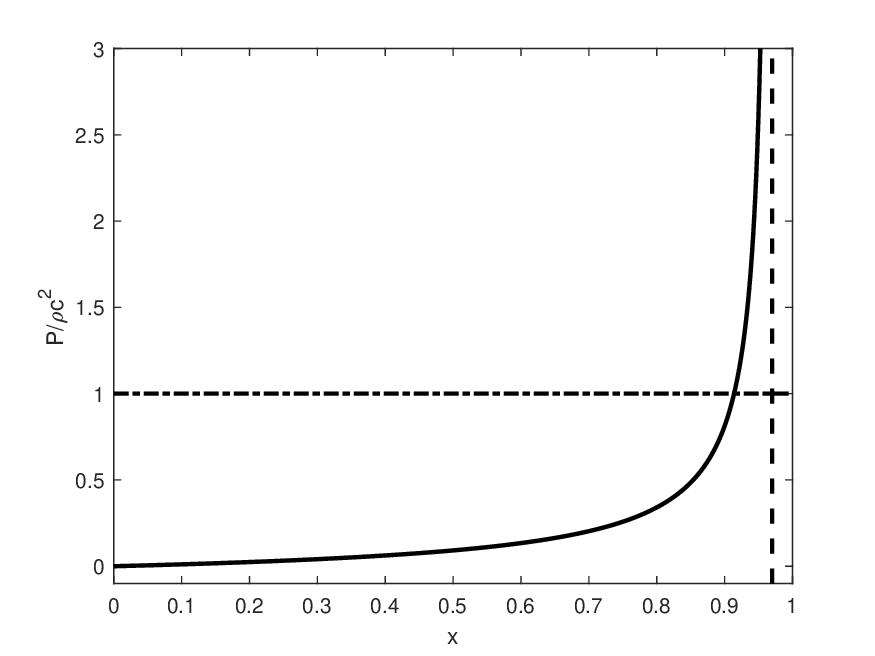}
\end{center}
\caption{(Solid line) Dependence of $\frac{P}{\rho c^2}$  on the parameter x , according to (\ref{eq43}). The abscissa of the vertical dashed line $x = x_l$
corresponds to the zero root of the denominator of  (\ref{eq43}), $x_l = 0.9849$ . The horizontal dash-dotted line $\frac{P}{\rho c^2}=1$ separates
the physically admissible region from the upper region where the principle of causality is violated.}
\label{1}
\end{figure}

As the denominator on the right-hand side of (\ref{eq43})
decreases to zero, the ratio $\frac{P}{\rho c^2}$ and the density $\rho$ tend
to infinity. The abscissa of the vertical dashed line $x=x_l$
in Fig. \ref{1} corresponds to the zero root of the
denominator of (\ref{eq43}), $x_l=0.9849$. The horizontal
dash-dotted line $\frac{P}{\rho c^2}=1$ separates the physically permissible
region from the upper region, where the principle
of causality is violated, according to which the
speed of sound in matter cannot exceed the speed of
light. As follows from rigorous calculations based on
the numerical solution of the Oppenheimer–Volkoff
differential equilibrium equation \cite{bk89,zn71}, NSs with different
equations of state become unstable at significantly
lower values of the parameter $x=\frac{R_g}{R}\sim \frac{1}{3}$ ,
which can be slightly larger or slightly smaller than 1/3
for different equations of state. The corresponding
parameters of NSs within the uniform density model
are calculated in the subsequent sections.

\noindent

The asymptotic solution of (\ref{eq43}) at $x\rightarrow x_l$ is
obtained analytically using the decompositions

 \begin{eqnarray}
\label{eq54}
x=x_l-\delta, \quad \Phi_1(x)=\Phi_1(x_l)-\Phi^{'}_1(x_l)\delta.
\end{eqnarray}
Since $\Phi_1(x_l)=0$ , the asymptotic solution of (\ref{eq43})
has the form

 \begin{eqnarray}
\label{eq55}
\frac{P}{\rho c^2}\simeq\frac{\Phi_0(x_l)}{-\Phi^{'}_1(x_l)\delta}\simeq\frac{0.02493}{\delta},\qquad\qquad\qquad \nonumber\\
\Phi_0(x_l)=x_l(3-x_l^2)-3\sqrt{1-x_l^2}\,\sin^{-1}x_l\simeq 1.275,\\
\Phi{'}_1(x_l)=
 21x_l^2-9 \frac{x_l}{\sqrt{1-x_l^2}}\,\sin^{-1}x_l\simeq-51.15 . \nonumber
\end{eqnarray}
This asymptotics is shown in Fig. \ref{1}.

\begin{figure}[h!]
\begin{center}
\includegraphics[width=0.7\linewidth]{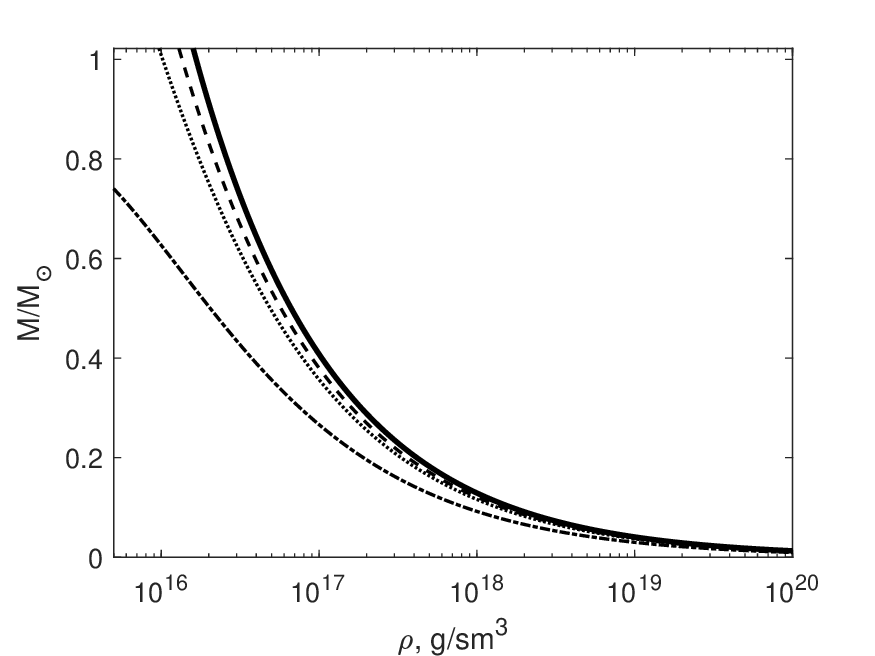}
\end{center}
\caption{ Dependence $M(\rho)$ for different equations of state. The solid line shows asymptotics (\ref{eq57}). The dashed line shows the
decrease of $M(\rho)$ in model I H, the dotted line refers to a degenerate neutron gas with a quadratic correction, and the dash-dotted
line corresponds to a degenerate neutron gas.}
\label{2}
\end{figure}

 Interestingly, in a uniform-density star, the dependence $M(\rho)$
for unstable models near the threshold $x=x_l$
is universal and does not depend on the equation
of state. Indeed, near the boundary $x=x_l$, the
relations from (\ref{eq41}) for $M$ and $x^2$ are written as

 \begin{eqnarray}
\label{eq56}
M=\frac{4\pi}{3}\rho R^3,\quad x^2=x_l^2=0.9845^2=0.97=\frac{2GM}{c^2 R}, \quad M=0.97\frac{c^2 R}{2G}.
\end{eqnarray}
Hence, from two expressions for $M$, we obtain asymptotic
dependences $R(\rho)$ and $M(\rho)$ in the form

\begin{eqnarray}
\label{eq57}
R= \left(\frac{0.97 c^2}{2G \rho}\right)^{1/2}
\sqrt{\frac{3}{4\pi}}, \qquad
 M=\sqrt{\frac{3}{4\pi}} \left(\frac{0.97 c^2}{2G}\right)^{3/2}
 \frac{1}{\sqrt{\rho}}.
\end{eqnarray}
The asymptotic dependence $M(\rho)$, the same for all
equations of state, is shown in Fig. \ref{2}.

The solutions near the limiting value do not have
any special physical meaning, since they are obtained
from an approximate uniform density model, which is
generally inapplicable for unstable star models at densities
significantly exceeding the maximum mass density,
when the curve $M(\rho)$ is qualitatively different
from exact solutions with an arbitrary number of
degrees of freedom \cite{dh},\cite{htww}. It is worth noting the possibility
of using the Galerkin method to obtain equilibrium
solutions \cite{bkd98} when the density is represented as
a sum over given functions with coefficients that minimize
the energy functional. This will lead to an
increase in the number of degrees of freedom and to
obtaining a more accurate solution, compared to a
uniform distribution, which is determined by a system of algebraic equations. The increase in the number of
functions will lead to a qualitative improvement of the
solution at very high densities. With an increase in the
number of functions, the density at which the solution
is qualitatively reliable increases. The dependence $M(\rho)$
obtained is consistent with the conclusion of
\cite{zeld1} about the metastability of any mass relative to relativistic
collapse and that, for any gravitating mass,
there are at least two equilibrium states, of which only
the state with the lowest density is stable. These states
are separated by a large potential barrier with a negligible
but finite probability of penetration through it
and subsequent relativistic collapse.

 It is believed that, for realistic star models, the
principle of causality is satisfied, according to which
the speed of sound cannot exceed the speed of light.
To satisfy this condition, the inequality $P\le \rho c^2$must
be satisfied \cite{zeld2}, which is determined by the horizontal
line with the ordinate $y=1$ in Fig. \ref{1}, as well as vertical
lines in subsequent figures, corresponding to densities
where $P=\rho c^2$. The pressure in the asymptotic models
is determined by the equation of state using (\ref{eq55}).

\section{ ALGORITHM FOR CONSTRUCTING
AN EQUILIBRIUM CURVE $M(\rho)$
FOR A UNIFORM DENSITY MODEL
OF A STAR WITH A GIVEN EQUATION
OF STATE}
The dependences $M(\rho)$ or $M(\rho_0)$ and $M(R)$ can be
built using the following procedure.

1. Select an equation of state $P(\rho)$ or $P(\rho_0)$ given by
formulas or tables.

2. Set the ratio of the gravitational radius of the star
to the physical radius: $x^2= \frac{R_g}{R}$.

3. Find from (\ref{eq43}) the ratio of pressure to density..

4. Using formulas or tables specifying the equation
of state, find the value of the density at which relation
(\ref{eq43}) is satisfied.

5. Find the radius of the model from the definition
of $x$ in the form $R^2=\frac{3c^2 x^2}{8\pi G\rho}$.

6. Finally, find the mass of the uniform-density
model, $M=\frac{4\pi}{3} \rho R^3$.
The calculation results for several
equations of state are presented in Figs. \ref{2}–\ref{9}.

\section{EQUATIONS FOR DETERMINING
THE CRITICAL MASS OF A UNIFORM-DENSITY
NEUTRON STAR
WITH AN ARBITRARY EQUATION
OF STATE}
As shown by Ya.B. Zel’dovich \cite{zeld63} (see also  \cite{bk89,zn71,htww,zn65,sht85}), the hydrodynamic stability of a star is lost
at the point of the mass-density curve where the mass
has a maximum. In general relativity, the maxima of
the curves $M(\rho)$ and $M(\rho_0)$ coincide \cite{zn71}; therefore,
any of these curves can be used. Let us consider
the dependence $M(\rho)$. Given (\ref{eq43b}), Eq. (\ref{eq43}) can be
written as

    \begin{eqnarray}
\label{eq44a}
g(\rho)=\Phi(\rho,M),
\end{eqnarray}
where
 \begin{eqnarray}
g(\rho)=\frac{P}{\rho c^2},\quad 
\Phi(\rho,M)=\frac{\Phi_0(x)}{\Phi_1(x)},\quad \nonumber
x=R\sqrt{D}=\sqrt{\frac{R_g}{R}}=\sqrt{\frac{8\pi G}{3}M^{2/3}
 \rho^{1/3}},\quad R=\frac{3}{4\pi}\bigl(\frac{M}{\rho}\bigr)^{1/3}. \nonumber
\end{eqnarray}
Differentiating expression (\ref{eq44a}), we obtain

\begin{eqnarray}
\label{eq45a}
\frac{d g(\rho)}{d \rho}=\frac{\partial \Phi}{\partial M}
\frac{dM}{d\rho} + \frac{\partial \Phi}{\partial \rho}.
\end{eqnarray}
Considering that, at the maximum of the mass, $\frac{dM}{d\rho}=0$
, we obtain the following system of equations
for determining the parameters $M_{max}$ and $\rho_{cr}$ of the
critical state of a neutron star:

\begin{eqnarray}
\label{eq46a}
g(\rho)=\Phi(\rho,M), \qquad
\frac{d g(\rho)}{d \rho}=\frac{\partial \Phi}{\partial \rho}.
\end{eqnarray}
Here, the right-hand sides of Eqs. (\ref{eq46a}) have the same
form for all equations of state and are calculated by
formula (\ref{eq44a}). The left-hand sides of Eqs. (\ref{eq46a}) are
determined by the equation of state, specified by an
analytical formula or from a table.

\section{CALCULATIONS AND RESULTS}
To build solutions of the algebraic equation for various
equations, we used the system of equations (\ref{eq41})
and (\ref{eq43}). The solution was carried out by Newton’s
method \cite{ns} with a relative accuracy of $10^{-7}$ . Models
of cold NSs were constructed using the equation of
state of a degenerate neutron gas and with more realistic
equations of state from \cite{bethe1,bethe}. Analytical expressions
for $P$ and $\varepsilon$ (total energy density) in models 3–
6 (see below) were obtained as a result of approximation
of tabulated data in \cite{bethe}.
  \\

\begin{figure}[!h]
\begin{center}
\includegraphics[width=0.65\linewidth]{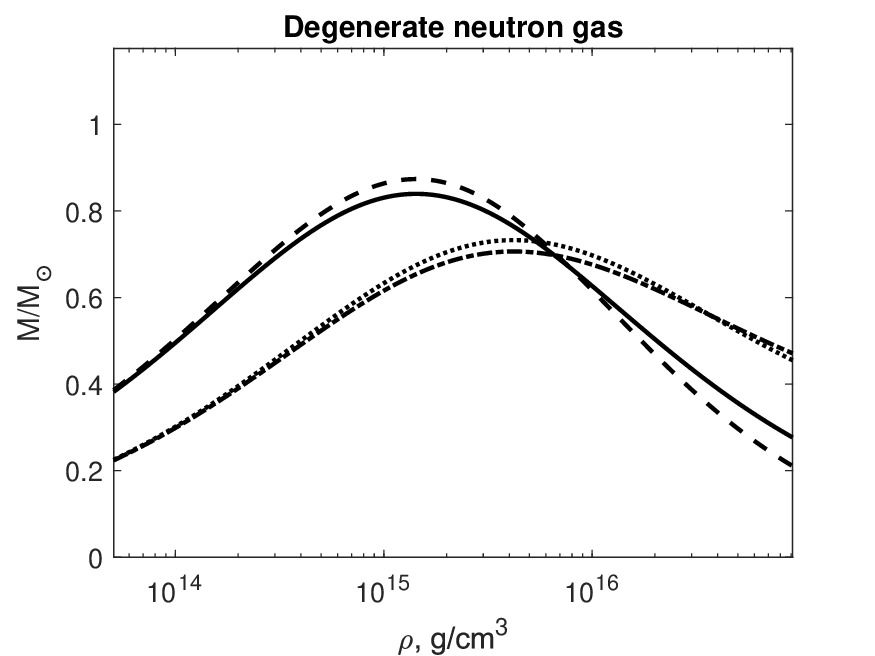}
\end{center}
\caption{ Degenerate neutron gas: dependence $M(\rho)$ in the case of (solid line) a uniform density distribution and (dash-dotted line) the exact model; dependence $M_0(\rho)$ (dashed line) in the case of uniform density and (dotted line) the exact model.}
\label{3}
\end{figure}

{\bf 1.  Model of a degenerate neutron gas.} It is the simplest
case, which does not consider interactions \cite{oveq}
(see Fig. \ref{3}):

\begin{equation*}
     \begin{cases}
     P = \dfrac{m^4c^5}{24\pi^2\hbar^3} (y(2y^2-3)\sqrt{1+y^2}+3\sinh^{-1}y) = 6.859 \times 10^{35}  (y(2y^2-3)\sqrt{1+y^2}+3\sinh^{-1}y) \quad dyn/cm^2,\\[7pt]
     \varepsilon = \dfrac{m^4c^5}{24\pi^2\hbar^3 } (3y(2y^2+1)\sqrt{1+y^2}-3\sinh^{-1}y)
     =6.859\times10^{35} (3y(2y^2+1)\sqrt{1+y^2}-3\sinh^{-1}y) \quad erg/cm^3,\\[9pt]
     y = \left(\dfrac{3\pi^2\rho_0}{m}\right)^{1/3} \frac{\hbar}{mc} =  \left(\dfrac{\rho_0}{6.1\times10^{15}\quad g/cm^3}\right)^{1/3}\\[8  pt]
     \end{cases}
\end{equation*}

\begin{figure}[h!]
\begin{center}
\includegraphics[width=0.65\linewidth]{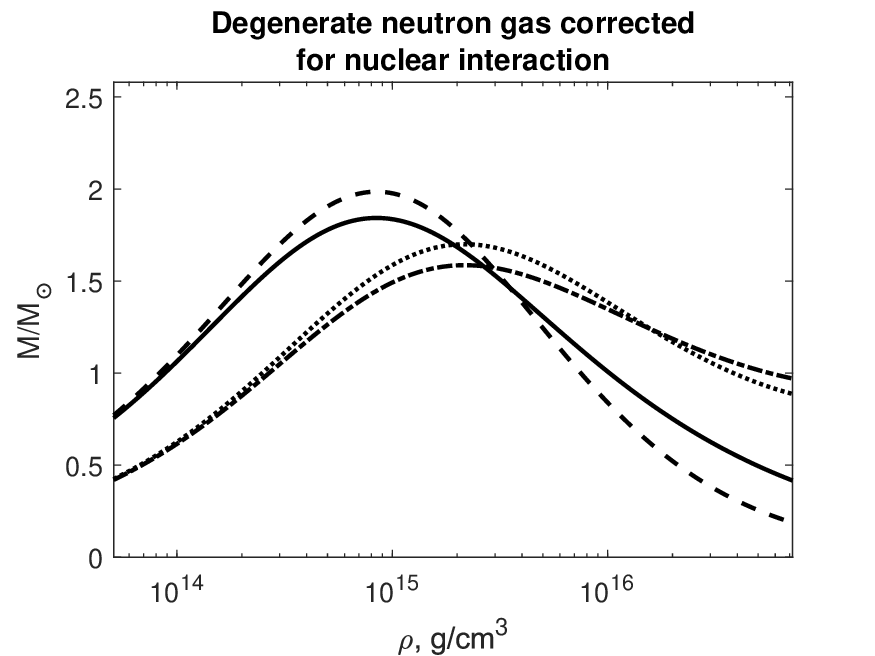}
\end{center}
\caption{ Degenerate neutron gas with a correction for nuclear interaction: the dependence $M(\rho)$ (solid line) in the case of a uniform
density distribution and (dash-dotted line) the exact model; the dependence $M_0(\rho)$ (dashed line) in the case of uniform density
and (dotted line) in the exact model.}
\label{4}
\end{figure}

{\bf 2.  Model of a degenerate neutron gas with a quadratic
correction for density.} This equation of state
approximately considers nuclear interaction \cite{bk68} (see
Fig. \ref{4}):

\begin{equation*}
     \begin{cases}
     P = \dfrac{m^4c^5}{24\pi^2\hbar^3} (y(2y^2-3)\sqrt{1+y^2}+3\sinh^{-1}y) + \dfrac{6\pi\hbar^3}{m_p^4 c} \rho_0^2 = \\ =6.859 \times 10^{35}(y(2y^2-3)\sqrt{1+y^2}+3\sinh^{-1}y) + 0.9421086\cdot10^5\rho_0^2 \quad dyn/cm^2,\\[7pt]
     \varepsilon = \dfrac{m^4c^5}{24\pi^2\hbar^3 } (3y(2y^2+1)\sqrt{1+y^2}-3\sinh^{-1}y)+\dfrac{6\pi\hbar^3}{m_p^4 c} \rho_0^2 =\\[7pt]
     =6.859\times10^{35}(3y(2y^2+1)\sqrt{1+y^2}-3\sinh^{-1}y)+0.9421086\times10^5\rho_0^2 \quad erg/cm^3\\[9pt]

     \end{cases}
\end{equation*}

\begin{figure}[!h]
\begin{center}
\includegraphics[width=0.65\linewidth]{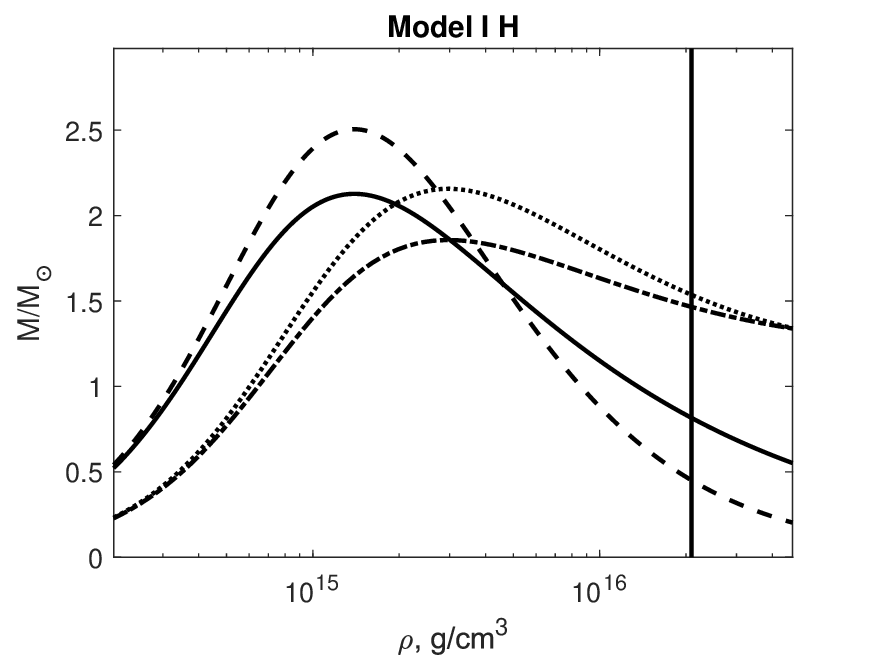}
\end{center}
\caption{ Model I H. The dependence $M(\rho)$ (solid line) in the case of a uniform density distribution and (dash-dotted line) in the
exact model; the dependence $M_0(\rho)$ (dashed line) in the case of a uniform density and (dotted line) in the exact model. The vertical
solid line shows at what density $v_s=c$ and $\rho = 2.2 \times10^{16}$ g/cm$^3$.}
\label{5}
\end{figure}

{3. \bf Model I H, based on the Reid potential, the same
for all baryons} (see Fig. \ref{5}):
\begin{equation*}
    \begin{cases}
P = 586  \left(\frac{\rho}{m_n \times 10^{39}}\right)^{2.48}\times 10^{33}\, 
 \quad dyn/cm^2,\\
\varepsilon = n (15.05 + 3.96n^{1.48})\times 10^{35}\,    
   \quad erg/cm^3,\\[7pt]
    n = \dfrac{\rho_0}{m_n}
    \end{cases}
\end{equation*}

\begin{figure}[!h]
\begin{center}
\includegraphics[width=0.65\linewidth]{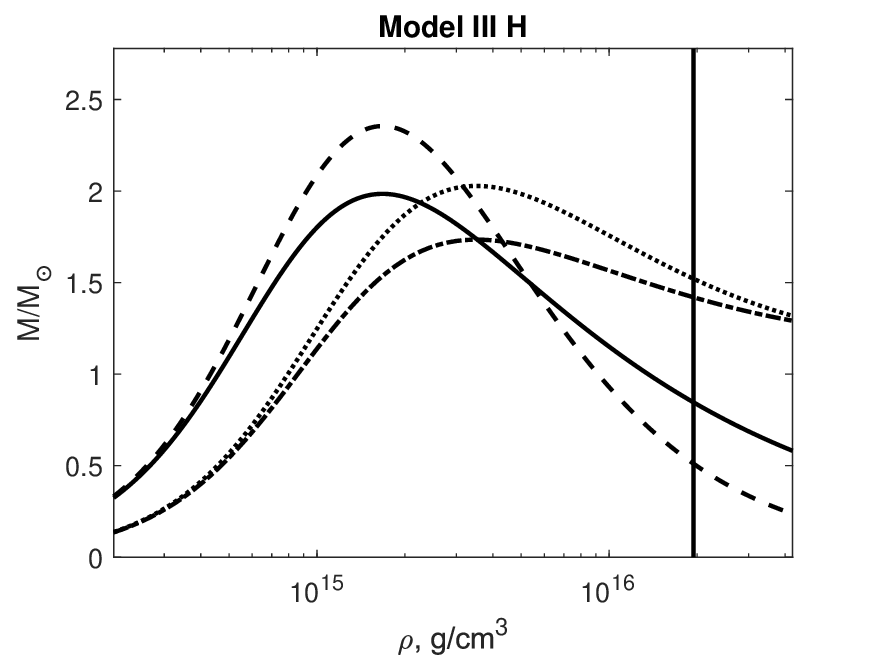}
\end{center}
\caption{ Model III H. The dependence $M(\rho)$ (solid line) in the case of a uniform density distribution and (dash-dotted line) in the
exact model; the dependence $M_0(\rho)$ (dashed line) in the case of a uniform density and (dotted line) in the exact model. The vertical
solid line shows at what density $v_s=c$ and $\rho = 1.91 \times10^{16}$ g/cm$^3$.}
\label{6}
\end{figure}

{\bf 4. Model III H, based on a more realistic potential
for np-interaction} (see Fig. \ref{6}):
\begin{equation*}
    \begin{cases}
P = 474 \left(\frac{\rho}{m_n \times 10^{39}}\right)^{2.55}\times 10^{33}   
     \quad dyn/cm^2,\\
\varepsilon = n (15.05+3.06n^{1.55})\times 10^{35}    
    \quad erg/cm^3\\[7pt]
   
    \end{cases}
\end{equation*}

\newpage

\begin{figure}[!h]
\begin{center}
\includegraphics[width=0.65\linewidth]{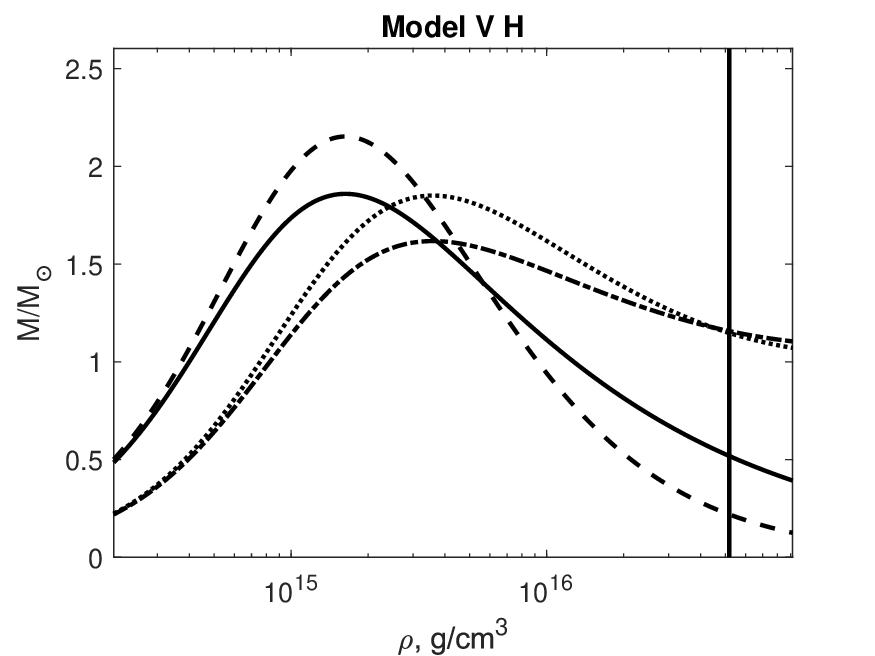}
\end{center}
\caption{ Model V H. The dependence $M(\rho)$ (solid line) in the case of a uniform density distribution and (dash-dotted line) in the
exact model; the dependence $M_0(\rho)$ (dashed line) in the case of a uniform density and (dotted line) in the exact model. The vertical
solid line shows at what density $v_s=c$ and $\rho =  2.15 \times10^{16}$ g/cm$^3$. }
\label{7}
\end{figure}

{\bf 5. Model V H, similar to model V N (see below), but
considering the creation of hyperons} (see Fig. \ref{7}):
\begin{equation*}
    \begin{cases}
P = 403  \left(\frac{\rho}{m_n \times 10^{39}}\right)^{2.33}\times 10^{33}   
   \quad dyn/cm^2,\\
\varepsilon = n (15.05+3.03n^{1.33   })\times 10^{35}    
    \quad erg/cm^3\\[7pt]
   
    \end{cases}
\end{equation*}

\begin{figure}[!h]
\begin{center}
\includegraphics[width=0.65\linewidth]{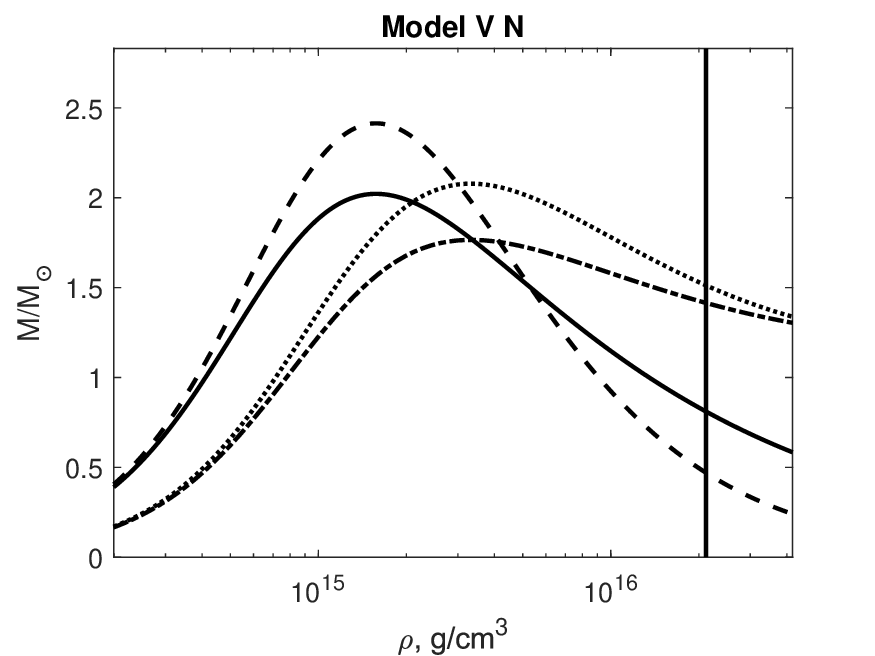}
\end{center}
\caption{Model V N. The dependence $M(\rho)$ (solid line) in the case of a uniform density distribution and (dash-dotted line) in the
exact model; the dependence $M_0(\rho)$ (dashed line) in the case of a uniform density and (dotted line) in the exact model. The vertical
solid line shows at what density $v_s=c$ and $\rho =  6.64 \times10^{16}$ g/cm$^3$. }
\label{8}
\end{figure}

{\bf  6. Model V N. Model of nucleon interaction considering
experimental data on the $\omega$-meson creation at
high energies} (see Fig. \ref{8}):

\begin{equation*}
    \begin{cases}
 P = 490 \left(\frac{\rho}{m_n \times 10^{39}}\right)^{2.508}\times 10^{33}   
   \quad dyn/cm^2,\\
\varepsilon = n (14.89+3.25n^{1.508})\times 10^{35}    
    \quad erg/cm^3\\[7pt]
    
    \end{cases}
\end{equation*}

\begin{figure}[!h]
\begin{center}
\includegraphics[width=0.65\linewidth]{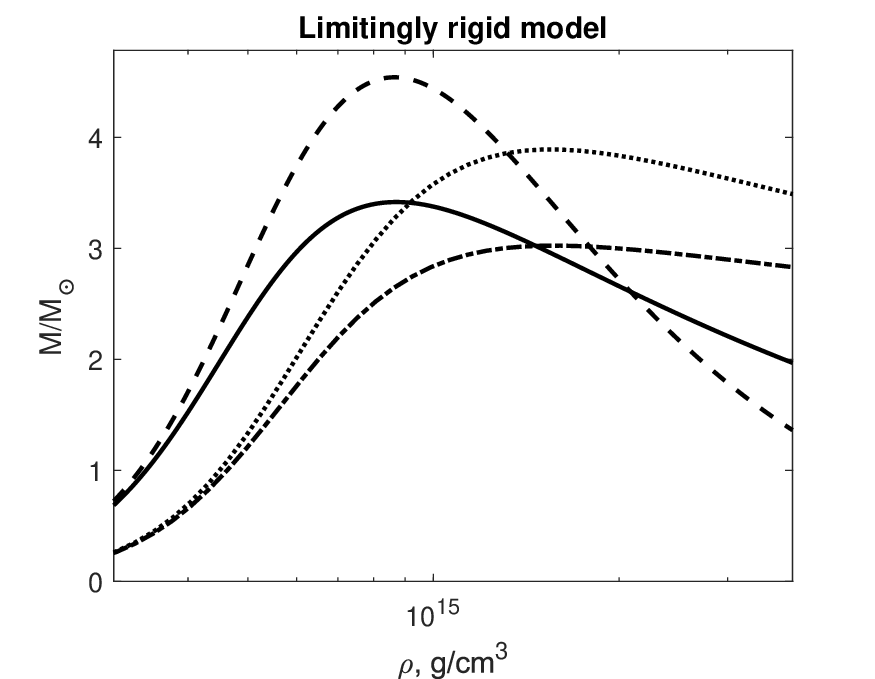}
\end{center}
\caption{Limitingly stiff model. The dependence $M(\rho)$ (solid line) in the case of a uniform density distribution and (dash-dotted
line) in the exact model; the dependence $M_0(\rho)$ (dashed line) in the case of a uniform density and (dotted line) in the exact
model.   }
\label{9}
\end{figure}

{\bf 7. Limitingly stiff equation of state} (see Fig. \ref{9}):
\begin{equation*}
    P = P^{*} + (\varepsilon-\varepsilon^{*})
\end{equation*}

To confirm the correctness of the calculations and
compare uniform- and nonuniform-density NS models,
we constructed models of cold NSs in the case of a
nonuniform density distribution. To find models with
different equations of state, it is necessary to integrate
the Oppenheimer–Volkoff equations with the thermodynamic
functions 1–7 given above:

\begin{equation*}
    \begin{cases}
    \dfrac{dP}{dr} = -\dfrac{G}{r^2c^2}\dfrac{(\varepsilon+P)(m+\dfrac{4\pi r^3P}{c^2})}{1-\dfrac{2Gm}{rc^2}}\\[15pt]
    \dfrac{dm}{dr} = 4\pi r^2 \dfrac{\varepsilon}{c^2}
     \end{cases}
\end{equation*}
Integration was carried out using the 4th-order
Runge–Kutta method with automatic step selection
\cite{ns}. In addition to building the dependencies M($\rho$)
shown in Fig. \ref{3}-\ref{9}, the dependences of rest mass on density,$M_0$($\rho$), were plotted in cases of uniform and nonuniform
density distribution. For uniform-density
NSs, this dependence is constructed relatively simply,
using Eqs. (\ref{eq20}) (see Figs. \ref{3}-\ref{9}). In the case of a nonuniform
density model, it is necessary to integrate the following
equation:

\begin{equation*}
    m_0 = 4\pi \int_0^r \dfrac{\rho r'^2}{\sqrt{1-\dfrac{2Gm'}{r'c^2}}}dr';\quad M_0 = m_0(R);
\end{equation*}
The integration was also carried out using the
4th order Runge–Kutta method. The parameters of the critical state of an NS are given in Table 1. The
density values in the case of a nonuniform density
model, within the limits of error, coincide with the
values given in \cite{bethe}.

\begin{table}[!h]
\caption{Critical parameters of neutron stars for different equations of state}

\begin{center}
\begin{tabular}{llllllll}
\multicolumn{8}{c}
{\textbf{Uniform density distribution   }}
\\\toprule\midrule
   & Neutron gas & \begin{tabular}[c]{@{}l@{}}Neutron gas with\\  a quadratic correction\\ for density $\rho^2$\end{tabular} & I H & III H & V H & V N & \begin{tabular}[c]{@{}l@{}}Limitingly \\ rigid model\end{tabular}\\\hline\addlinespace
$\rho_0, 10^{15}\frac{\mbox{g}}{\mbox{cm}^3}$   & 1.3    & 0.74   & 1.2   & 1.44     & 1.4    &1.37  & 0.76 \\\addlinespace
$\rho, 10^{15}\frac{\mbox{g}}{\mbox{cm}^3}$     & 1.43     & 0.84   & 1.39    & 1.68     & 1.63 &1.57 & 0.87  \\\addlinespace
$M/M_\odot$                       & 0.84    & 1.84    & 2.13    & 1.98     & 1.86    & 2.02 & 3.42   \\\addlinespace
$M_0/M_\odot$                     & 0.87    & 1.99    & 2.5    & 2.36    & 2.15    & 2.41 & 4.54   \\\addlinespace
$R, \mbox{km}$                    & 6.54     & 10.1    & 8.99    & 8.27     & 8.16     & 8.49  & 12.3
\\\bottomrule
\multicolumn{8}{c}{}\\\addlinespace
\multicolumn{8}{c}{\textbf{Exact model}}
\\\toprule\midrule
   & Neutron gas & \begin{tabular}[c]{@{}l@{}}Neutron gas with\\  a quadratic correction\\ for density $\rho^2$\end{tabular} & I H & III H & V H & V N & \begin{tabular}[c]{@{}l@{}}Limitingly \\ rigid model\end{tabular}\\\hline\addlinespace
$\rho_{c0}, 10^{15}\frac{\mbox{g}}{\mbox{cm}^3}$   & 3.54         & 1.68        & 2.15    & 2.54      & 2.62    &2.43   & 1.07  \\\addlinespace
$\rho_c, 10^{15}\frac{\mbox{g}}{\mbox{cm}^3}$     & 4.2     & 2.17        & 2.97    & 3.52     & 3.59    & 3.33 & 1.57  \\\addlinespace
$M/M_\odot$                       & 0.706    & 1.59        & 1.86    & 1.73     & 1.62    & 1.77  & 3.02   \\\addlinespace
$M_0/M_\odot$    &0.733                      & 1.7        & 2.16        & 2.03         & 1.85        & 2.08     & 3.9       \\\addlinespace
$R, \mbox{km}$                    & 9.14     & 12.8        & 9.9    & 9.06     & 9.17    & 9.35 & 12.7
\\\bottomrule
\end{tabular}
\end{center}
\end{table}

\section{CONCLUSIONS}
Within the framework of exact general relativity, we
have obtained algebraic equation (\ref{eq43}), which makes it
possible to approximately determine the parameters of
a superdense star (neutron or quark) in a uniform density
model. The equation has a form that is universal
for any equation of state $P(\rho)$, where $\rho$ considers the
contribution of all types of energy. It enables one to
find approximate dependences $M(\rho)$ and $R(\rho)$ for
equilibrium stars and to find the critical mass of a star,
which is the maximum allowable in equilibrium for a
given equation of state. After reaching the density corresponding
to this maximum mass on the $M(\rho)$ curve,
the star becomes unstable and must collapse with the
formation of a black hole. A comparison of our results
for uniform density models with exact solutions of differential
equations of equilibrium for various equations
of state from \cite{bethe} shows that the approximate
value of the critical mass of a star can exceed the exact
value by at most $\sim20\%$. Critical densities in uniform
density models turn out to be significantly lower than
the values of central densities in exact models (see
Table 1) and are comparable with the average densities
of these models.  

Note that the existence of the limiting mass of a
cold white dwarf was first discovered by Stoner \cite{stoner30} within the framework of an approximate Newtonian
uniform density model. The value of the mass
obtained by him was $\sim 20\%$ higher than the exact value
of this mass obtained later \cite{chandra31,landau32}.

\end{document}